\documentclass[]{spie}
\usepackage{amsmath,amsfonts,amssymb}
\usepackage{graphicx}
\usepackage{caption}
\usepackage{subcaption}
\usepackage[colorlinks=true, allcolors=blue]{hyperref}

\title{On-sky capabilities and performance of the Keck All Sky Precision Adaptive Optics system}

\author[a]{A. Surendran}
\author[a]{M. Service}
\author[a]{J. R. Delorme}
\author[b]{C. M. Correia}
\author[c]{M. A. Van Dam}
\author[a]{E. Marin}
\author[a]{A. Bouchez}
\author[a]{P. Wizinowich}
\author[a]{J. Chin}
\author[a]{S. Cetre}
\author[a,j]{S. Ragland}
\author[d]{U. Conod}
\author[a]{J. Taylor}
\author[a]{S. Lilley}
\author[a]{L. Gers}
\author[a]{E. Wetherell}
\author[e]{J. Lu}
\author[e]{N. Stiegler}
\author[e]{A. Pusack}
\author[f]{Z. Haggard}
\author[g]{W. Gauvin}
\author[h]{J. T. Bernard}
\author[i]{D. Pescoller}
\author[i]{R. Biasi}

\authorinfo{Further author information: (Send correspondence to Avinash Surendran)\\Avinash Surendran.: E-mail: asurendran@keck.hawaii.edu}

\affil[a]{W. M. Keck Observatory, Kamuela, HI, USA}
\affil[b]{Space ODT - Optical Deblurring Technologies, Porto, Portugal}
\affil[c]{Flat Wavefronts, Christchurch, New Zealand}
\affil[d]{Wakea Consulting, Grenoble, France}
\affil[e]{University of California, Berkeley, CA, USA}
\affil[f]{University of California, Los Angeles, CA, USA}
\affil[g]{Swinburne University of Technology, Melbourne, Australia}
\affil[h]{Australian National University, Canberra, Australia}
\affil[i]{Microgate SRL, Bolzano, Italy}
\affil[j]{Steward Observatory, Arizona, USA}

\begin{document}
\maketitle

\begin{abstract}
The Keck All Sky Precision Adaptive optics (KAPA) project upgrades the Keck I adaptive optics system to enable laser tomography using a four laser guide star (LGS) asterism. KAPA is now in operation in both narrow field and wide field modes to optimize correction on-axis or over the science field of view of the camera. The use of four LGSs, in conjunction with a tomographic reconstructor and pseudo open-loop control, leads to a significant reduction in wavefront error. We describe the overall architecture, development of the tomographic algorithm, real-time implementation and preliminary on-sky results here. By comparing the on-sky image quality with that obtained using a single LGS (sLGS) we clearly demonstrate the benefits of laser tomography, a technology which is crucial to the success of the next generation of extremely large telescopes.
\end{abstract}


\section{Introduction}

\begin{figure}[ht]
\centering
\setlength{\fboxsep}{0.15pt}   % Distance between image and border
\setlength{\fboxrule}{0.15pt}  % Border line thickness
\fbox{\includegraphics[width=0.8\textwidth]{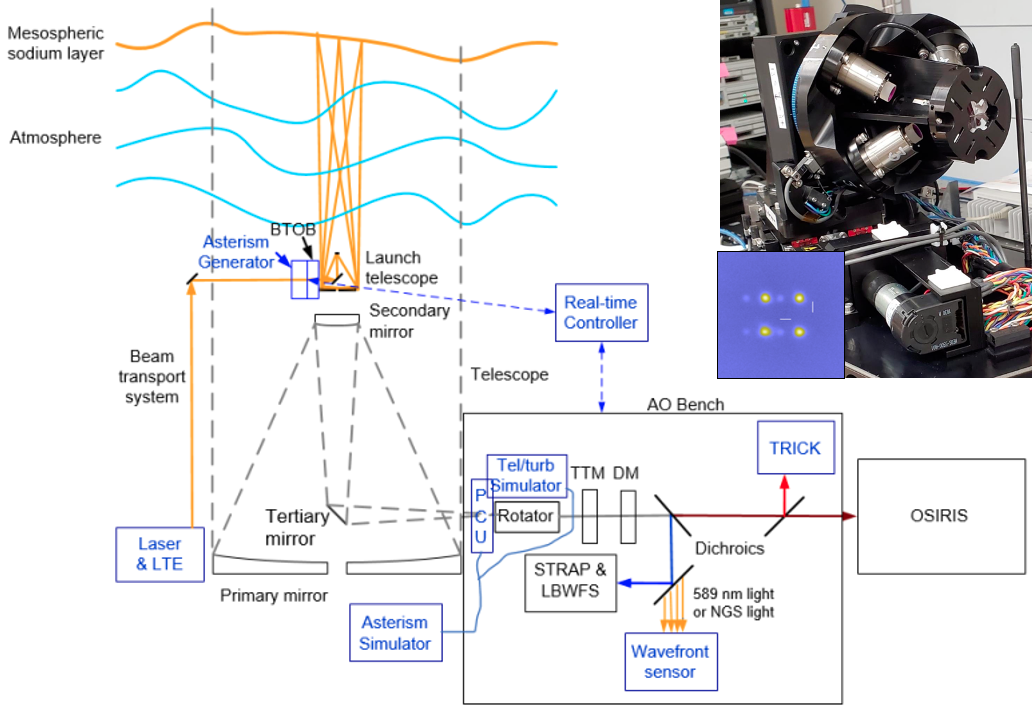}}
\caption{KAPA System Architecture which shows the additions for KAPA in blue. Inset shows a photo of the Asterism Generator and the four laser spots as seen on the AO acquisition camera.}
\label{fig:kapa-architecture}
\end{figure}

The current Keck I Adaptive Optics (AO) system\cite{Jason2012} can operate in either a single natural guide star (NGS) mode or a single laser guide star (sLGS) mode, using a 20 x 20 subaperture Shack-Hartmann wavefront sensor along with a visible or near infrared TT sensor (for LGS TT correction) and a 349-actuator deformable mirror (DM). The Keck All sky Precision AO (KAPA) project upgrades the Keck I AO system to improve corrected-field performance, increase sky coverage, and enable more quantitative science with OSIRIS\cite{larkin2006} through laser tomography and improves near-infrared low-order sensing \cite{Wizinowich20,Proceedings2022}. KAPA was developed to conduct four high impact science projects\cite{chu2022, wright2024, lu2020}: 
\begin{itemize}
    \item Constraining dark matter and dark energy via strong gravitational lensing.
    \item Testing General Relativity and studying supermassive black hole interactions at the Galactic Center.
    \item Characterizing galaxy kinematics and metallicity using rare highly magnified galaxies.
    \item Directly studying gas-giant protoplanets around the youngest stars.
\end{itemize}

KAPA is poised to convert the existing Keck I AO system into a laser tomographic AO system with changes that include the following\cite{Wizinowich20,Proceedings2022}: 
\begin{itemize}
    \item Replacing the existing laser on the Keck I Nasmyth platform with a TOPTICA/MPBC fiber laser on the elevation ring for sufficient sodium return.
    \item Implementing a rotating four-LGS asterism as part of the beam-train.
    \item Implementing a new wavefront sensor camera, specifically a lower noise, larger format OCAM2k camera with four sets of pupil relay optics.
    \item Implementing a new real-time controller to support laser tomography.
    \item Upgrading the existing controls and operations software to support all of the above changes.
\end{itemize}

This paper focuses on the first on-sky capabilities and preliminary measured performance of KAPA. The paper emphasizes commissioning outcomes such as the widening of the corrected field, the comparison between single LGS and laser tomographic AO (LTAO) on-axis performance and early science-field demonstrations including that of the GC.

\section{System Architecture}

Testing and calibration are supported by an asterism simulator source box in the AO electronics room that feeds five optical fibers (four 590 nm light sources simulating the 7.6 arcsec asterism and one broadband source simulating an on-axis natural guide star and science target) to the precision calibration unit (PCU) at the AO input focal plane \cite{Wizinowich20Daytime,freeman2023optical}. The real-time controller (RTC) interface module (IM) is located in the AO electronics room, with the GPU-based computational engine and the telemetry server in a computer room off the telescope \cite{CorreiaRTC}. The hardware sub-systems installed for KAPA are shown in Fig.~\ref{fig:kapa-architecture}.

\begin{figure}[ht]
\centering
\setlength{\fboxsep}{0.15pt}   % Distance between image and border
\setlength{\fboxrule}{0.15pt}  % Border line thickness
\fbox{\includegraphics[width=0.8\textwidth]{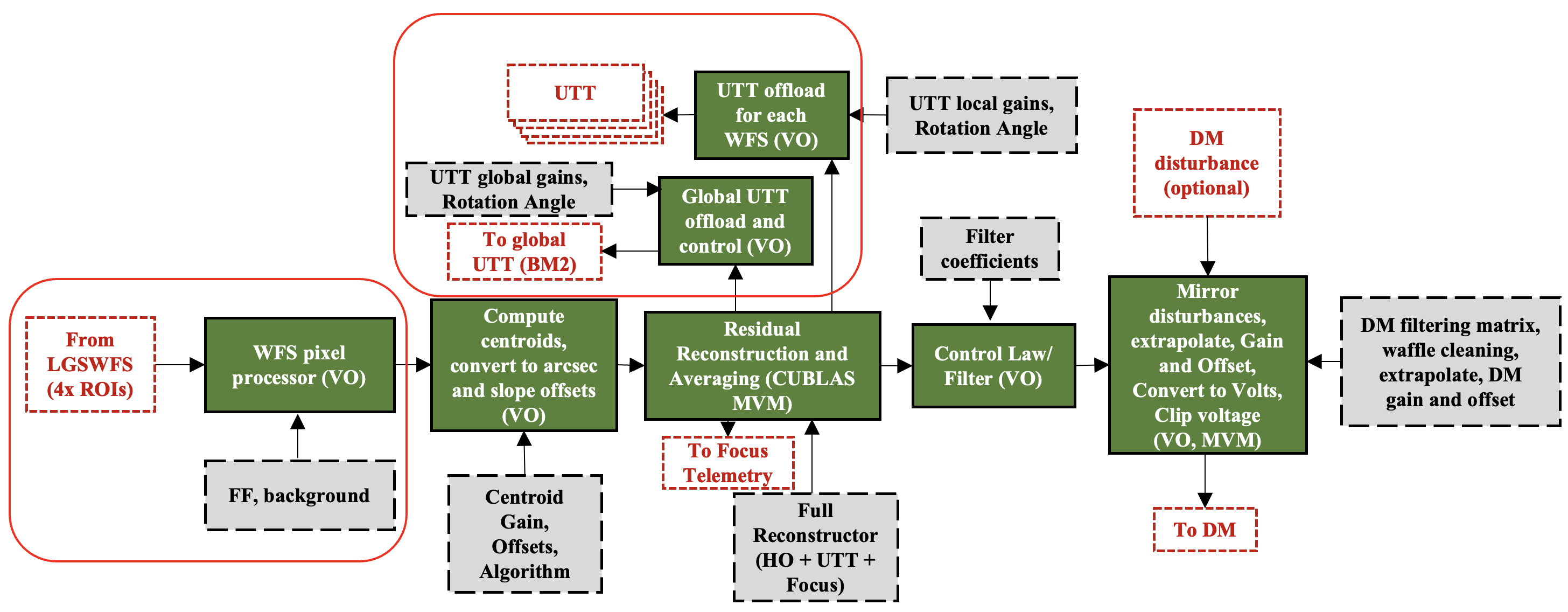}}
\caption{RTC control diagram showing multi-LGS averaging control (Phase 1) for KAPA. The green boxes are the compute blocks and the grey boxes are the configuration parameters set for each compute block. VO refers to a Vector Operation and MVM refers to a Matrix Vector Multiplication.}
\label{fig:kapa-glao-control}
\end{figure}

\begin{figure}[ht]
\centering
\setlength{\fboxsep}{0.15pt}   % Distance between image and border
\setlength{\fboxrule}{0.15pt}  % Border line thickness
\fbox{\includegraphics[width=0.6\textwidth]{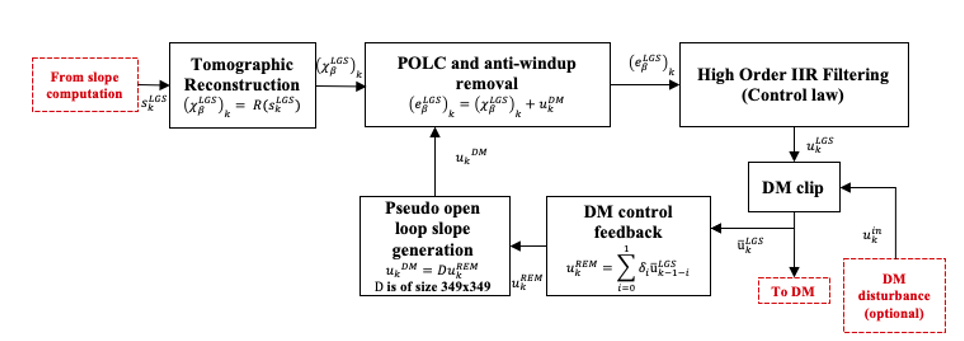}}
\caption{Tomographic control using a minimum variance LTAO reconstructor with POLC feedback (Phase 2)}
\label{fig:kapa-ltao-control}
\end{figure}

The tomographic control concept is organized into two phases. Phase 1 was focused on the multi-LGS averaging in a ground-layer-AO-like mode based on a simple averaging of four LGS phase outputs (Fig.~\ref{fig:kapa-glao-control}). The intent of this phase was for verification of multi-beacon sensing and multiple laser uplink tip-tilt (UTT) control functionality (red boxes in Fig.~\ref{fig:kapa-glao-control}). Phase 2 implements a minimum-variance LTAO reconstructor with pseudo open-loop (POLC) feedback (Fig.~\ref{fig:kapa-ltao-control}), using the four-beacon geometry and a covariance-based model of Maunakea turbulence\cite{Correia2014,Correia2015,vanDamKAON1519}. The equations that describe the realization of the complete tomographic reconstruction in Fig.~\ref{fig:kapa-ltao-control} are described in more detail in the Appendix (Section~\ref{sec:app} and Section~\ref{Sec:appendix_polc}).

\section{Commissioning and preliminary on-sky results}
\subsection{Daytime checks}

\begin{figure}[ht]
\centering
\setlength{\fboxsep}{0.15pt}   % Distance between image and border
\setlength{\fboxrule}{0.15pt}  % Border line thickness
\fbox{\includegraphics[width=0.5\textwidth]{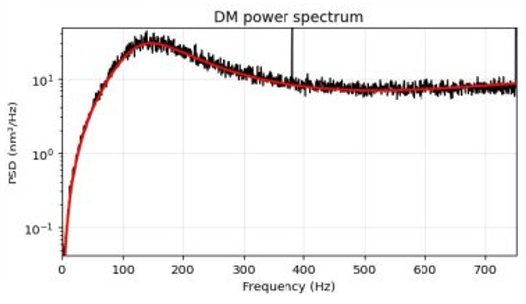}}
\caption{Residual power spectral density (closed loop response to measurement noise) at a WFS frame rate of 1.5 kHz with a DM loop gain of 0.5 for the multi-LGS averaging mode. The red curve represents the numerical model of the rejection transfer function (RTF).}
\label{fig:psd-kapa}
\end{figure}

Daytime calibrations (with the calibration light sources) were verified in October 2025 for four-pupil end-to-end closed-loop operation. This included verification of the control loop performance and calibration of non-common path aberrations (NCPA). The power spectral density (PSD) of the residual DM shape matched with a model corresponding to an end-to-end latency of 0.7 ms at a frame rate of 1.5 kHz for both Phase 1 (multi-LGS averaging) and Phase 2 (LTAO), similar to the latency observed with NGS and single-LGS modes \cite{Proceedings2022,KeckLGS2016}. An example of the data (black) and the fitted model (red) is shown in Fig~\ref{fig:psd-kapa} for the multi-LGS averaging configuration with the DM loop gain set to 0.5.

\subsection{Validation of Multi-LGS averaging mode}
\begin{figure}[ht]
\centering
\setlength{\fboxsep}{0.15pt}   % Distance between image and border
\setlength{\fboxrule}{0.15pt}  % Border line thickness
\fbox{\includegraphics[width=0.7\textwidth]{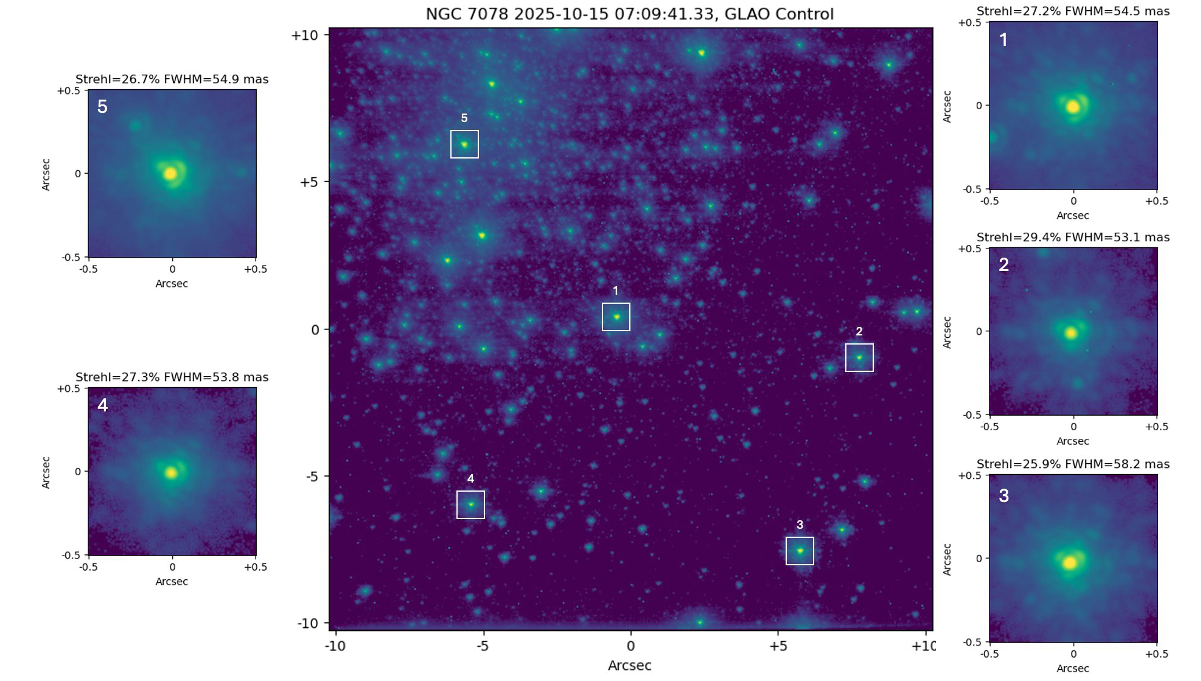}}
\caption{OSIRIS K-band Image of NGC 7078 using the multi-LGS averaging mode showing the uniformity of the PSF across the field}
\label{fig:gla-ngc7078}
\end{figure}

First on-sky tests with the multi-LGS averaging mode in October 2025 yielded promising results in reduction of angular anisoplanatism of the point spread function (PSF) across the field (Fig.~\ref{fig:gla-ngc7078}).

\subsection{LTAO First Light}

\begin{figure}
\centering
\begin{subfigure}{0.49\textwidth}
    \setlength{\fboxsep}{0.15pt}   % Distance between image and border
    \setlength{\fboxrule}{0.15pt}  % Border line thickness
    \fbox{\includegraphics[width=\textwidth]{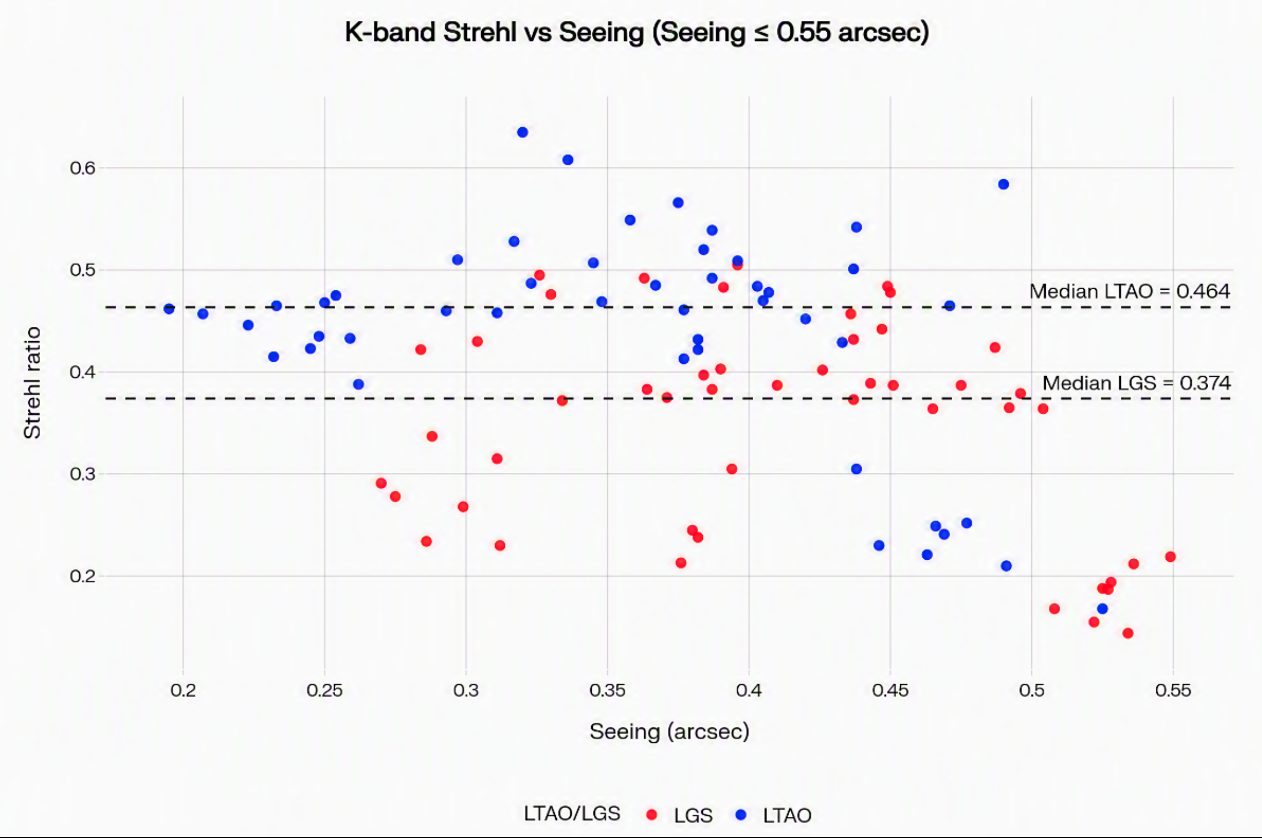}}
\end{subfigure}
\hfill
\begin{subfigure}{0.48\textwidth}
     \setlength{\fboxsep}{0.15pt}   % Distance between image and border
    \setlength{\fboxrule}{0.15pt}  % Border line thickness
    \fbox{\includegraphics[width=\textwidth]{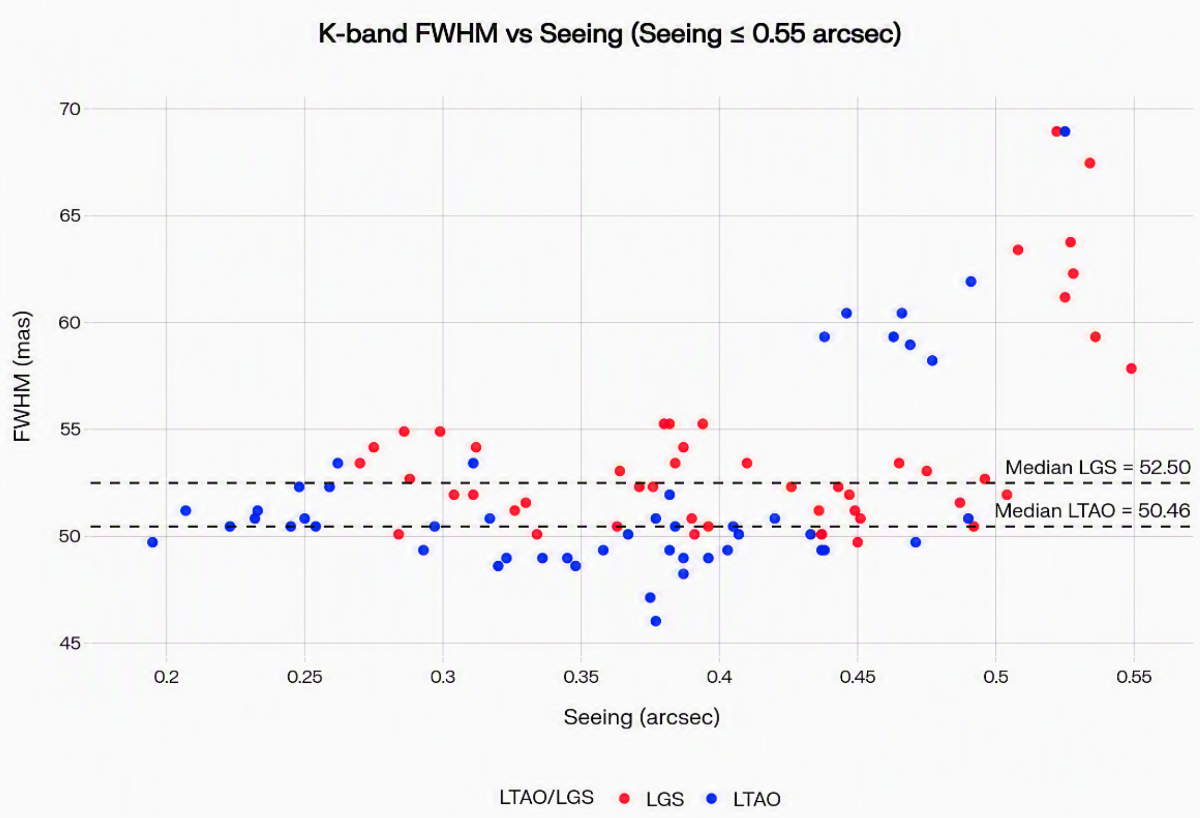}}
\end{subfigure}
        
\caption{SR and FWHM (mas) vs Seeing (arcsec) from the nights of Dec 4 and 6 for sLGS and LTAO modes.}
\label{fig:perf_vs_fwhm}
\end{figure}

\begin{figure}
\centering
\begin{subfigure}{0.49\textwidth}
    \setlength{\fboxsep}{0.15pt}   % Distance between image and border
    \setlength{\fboxrule}{0.15pt}  % Border line thickness
    \fbox{\includegraphics[width=\textwidth]{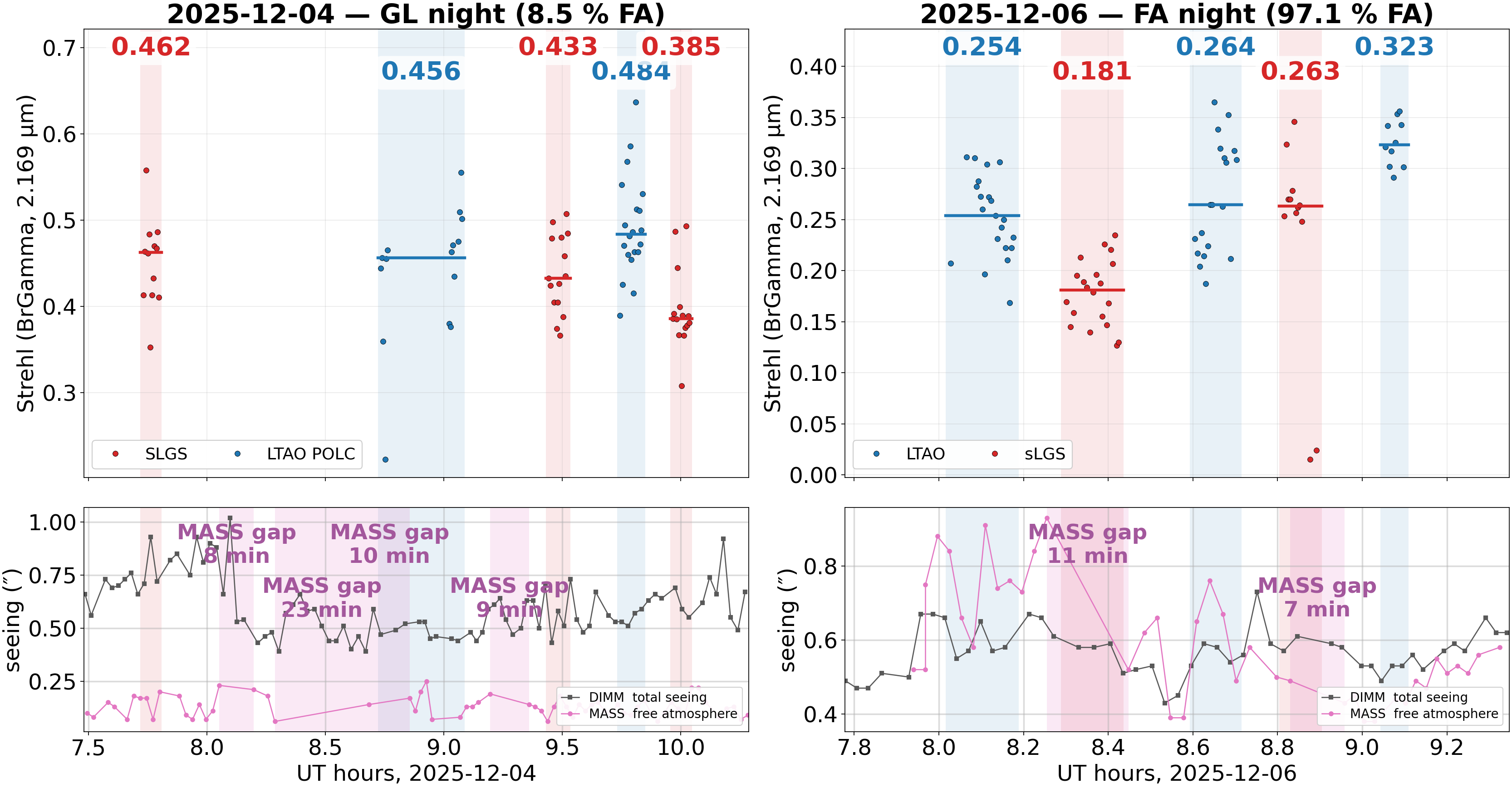}}
\end{subfigure}
\hfill
\begin{subfigure}{0.49\textwidth}
     \setlength{\fboxsep}{0.15pt}   % Distance between image and border
    \setlength{\fboxrule}{0.15pt}  % Border line thickness
    \fbox{\includegraphics[width=\textwidth]{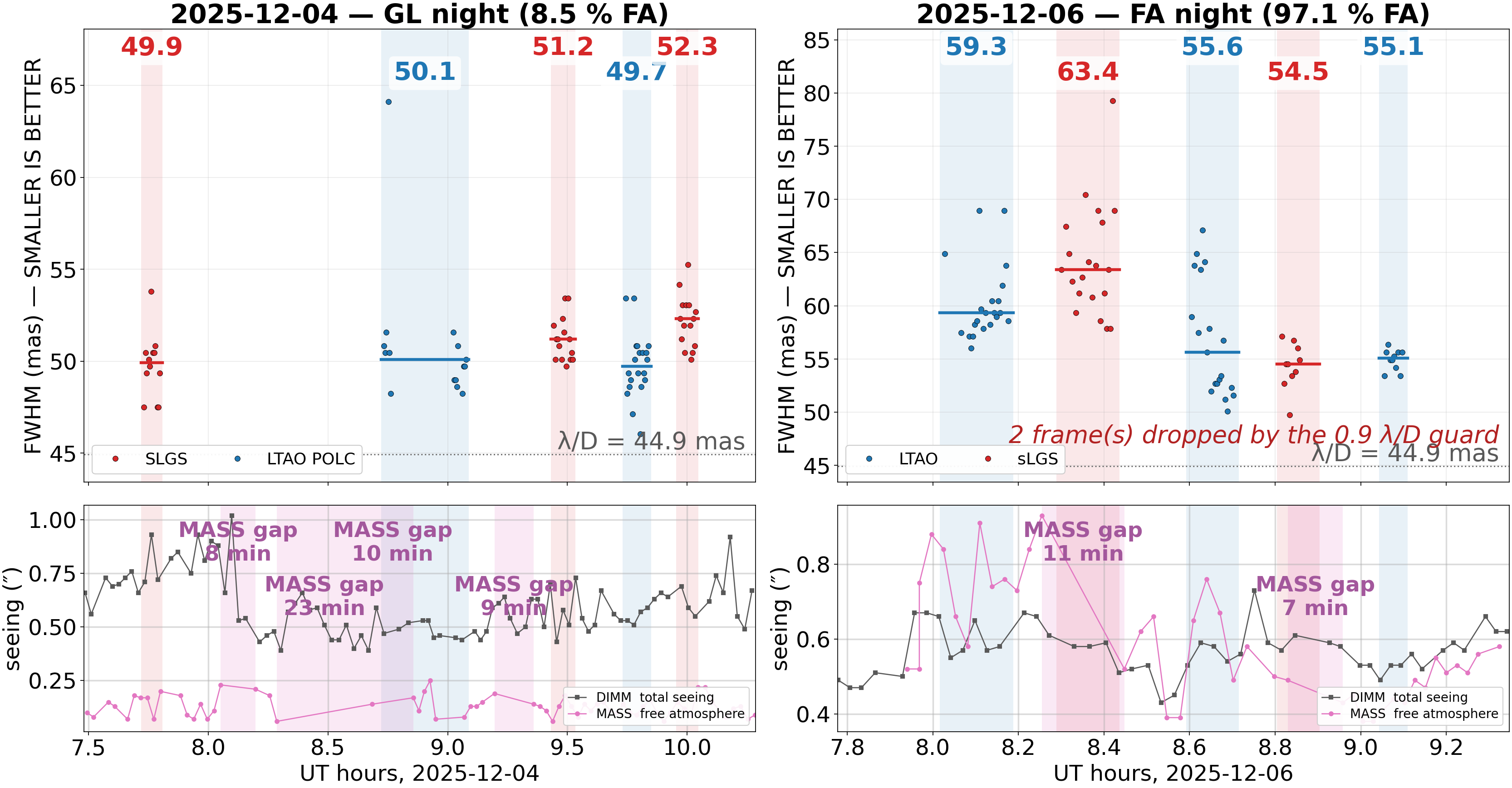}}
\end{subfigure}
\caption{SR and FWHM (mas) vs Seeing (arcsec) from the nights of Dec 4 and 6 for sLGS and LTAO modes, showing the comparison of LTAO performance improvement during a night dominated by free atmospheric seeing vs ground-layer seeing.}
\label{fig:SRFWHM_vs_MASSDIMM}
\end{figure}

First light with the full LTAO pipeline was achieved on Dec 4, 2025. After tuning the regularization and the reconstructor turbulence model, we compared sLGS vs LTAO performance through the nights of Dec 4 and 6, 2025. The improvement in Strehl Ratio (SR) and full-width-half-maximimum (FWHM) are shown in Fig.~\ref{fig:perf_vs_fwhm}. The seeing in the x-axis was extracted from fitting the RMS wavefront of the DM shape (obtained during the science exposure) to a Von-Karman profile to ensure on-axis seeing estimate (as opposed to reliance on the turbulence profile obtained from the Multi-Aperture Scintillation Sensor (MASS) and Differential Image Motion Monitor (DIMM) on the Maunakea weather center which could be pointed at a different direction towards the sky). This seeing is slightly underestimated due to the limited stroke and spatial frequency of the DM. All observations were performed with an on-axis TT star of R$\sim$12 and the science images were taken at a wavelength of 2.17 $\mu$m.

When we analyze the same set of observations (from December 4 and 6), we find predictably that the improvement in AO performance for LTAO as compared to sLGS is higher for the night of Dec 6 because of the presence of stronger free atmospheric seeing (and hence higher focal anisoplanatism) as compared to ground layer seeing. For the night of December 4, where ground layer seeing dominated the seeing profile, performance improvement is fairly modest for LTAO as compared to sLGS. The results are shown in Fig.~\ref{fig:SRFWHM_vs_MASSDIMM}.

\subsection{Performance maps on crowded fields}

\begin{figure}[ht]
\centering
\setlength{\fboxsep}{0.15pt}   % Distance between image and border
\setlength{\fboxrule}{0.15pt}  % Border line thickness
\fbox{\includegraphics[width=0.7\textwidth]{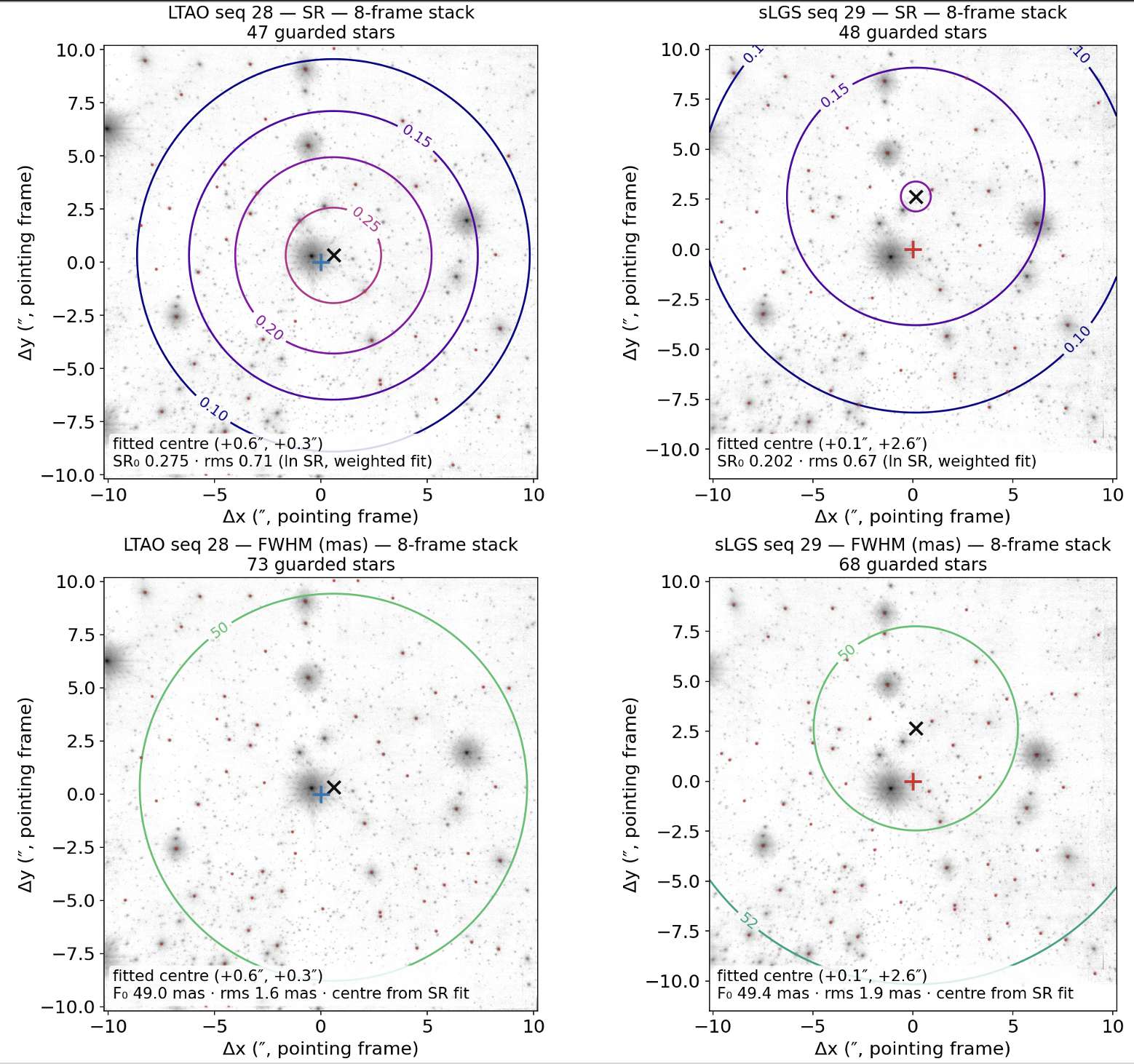}}
\caption{OSIRIS K-band image SR map (top row) and FWHM map (bottom row) of M79 in sLGS (right column) vs LTAO (left column). The black x is the estimated field position of best performance and the red and green crosses point to the field position that the telescope is pointed towards. Note that the SR map is modeled from different PSFs in the field and is not a true contour map.}
\label{fig:m79_onaxis_srmap}
\end{figure}

Fig.~\ref{fig:m79_onaxis_srmap} shows the degradation of SR and FWHM across the field compared between sLGS and LTAO for M79. As with earlier images, these images were taken in the K-band at similar MASS/DIMM seeing conditions. It is evident from the figure that LTAO provides a larger anisoplanatic angle (and hence a wider corrected field) than the sLGS mode in similar conditions, but we need to collect and analyze more data to quantify the improvement in the anisoplanatic angle in the LTAO mode.

\begin{figure}
\centering
\begin{subfigure}{0.49\textwidth}
    \setlength{\fboxsep}{0.15pt}   % Distance between image and border
    \setlength{\fboxrule}{0.15pt}  % Border line thickness
    \fbox{\includegraphics[width=\textwidth]{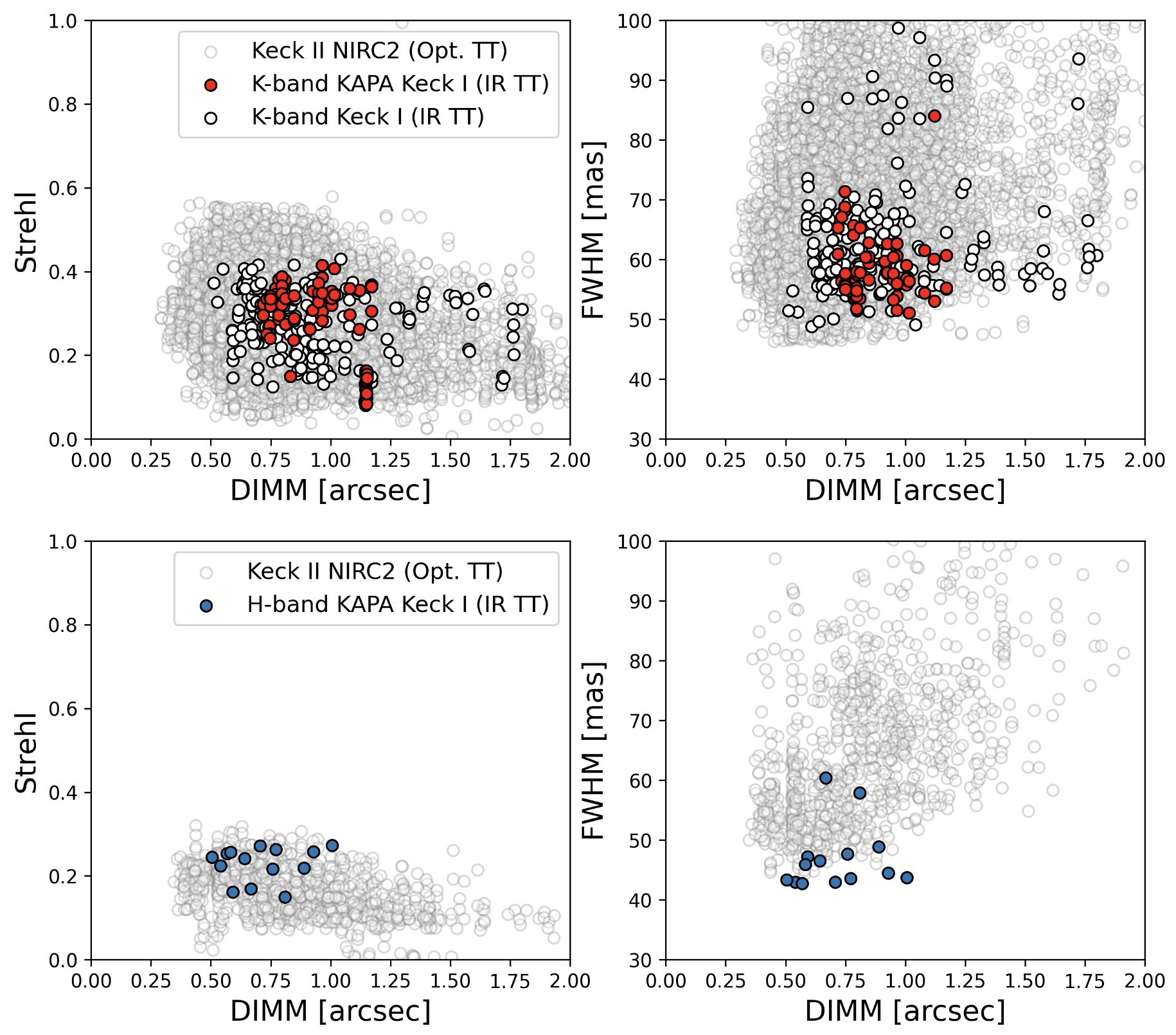}}
\end{subfigure}
\hfill
\begin{subfigure}{0.49\textwidth}
     \setlength{\fboxsep}{0.15pt}   % Distance between image and border
    \setlength{\fboxrule}{0.15pt}  % Border line thickness
    \fbox{\includegraphics[width=\textwidth]{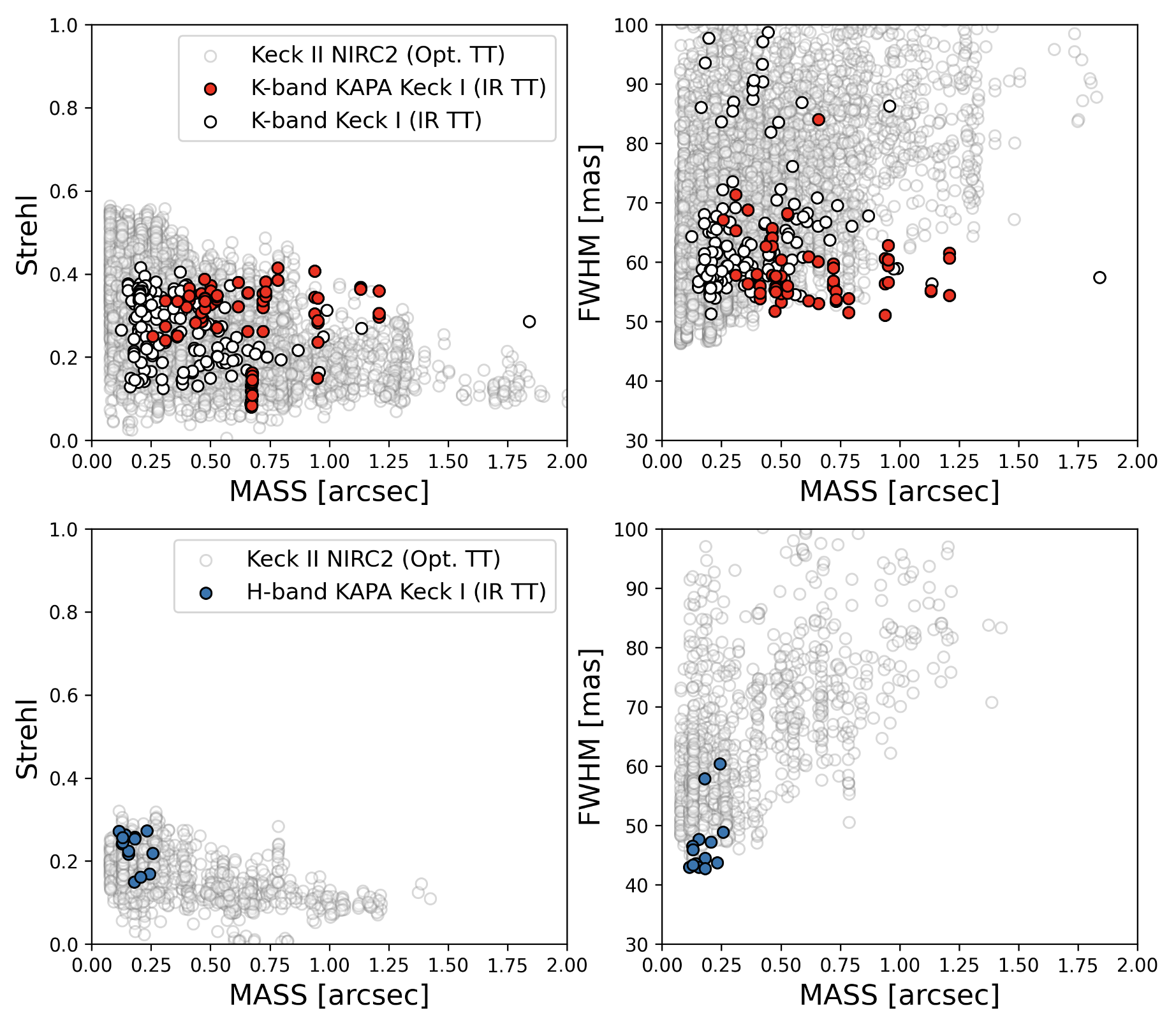}}
\end{subfigure}
\caption{Comparison of GC SR and FWHM in K and H bands compared to decades of image quality measurements on both OSIRIS and NIRC2 with optical TT correction on NIRC2 and NIR TT correction on OSIRIS. Most of the improvement in correction for KAPA in the image on the left (with respect to the integrated seeing observed by DIMM) could be attributed to the TT loop closure on IRS29N which is closer than IRS7. A higher turbulence on MASS means a stronger high altitude turbulence layer and the image on the right shows a considerable improvement in SR and FWHM in cases of strong MASS turbulence scenarios.}
\label{fig:GC_MASS_DIMM}
\end{figure}

Teams from the University of California Los Angeles (UCLA) and University of California Berkeley (UCB) have been observing the GC with OSIRIS on Keck I and NIRC2 on Keck II for more than a decade\cite{Dinh_2024,von_Fellenberg_2025,Yelda_2010} and this helps us compare the performance improvement provided by KAPA as compared to the baseline performance of the Keck I and II AO systems on the same target. Fig.~\ref{fig:GC_MASS_DIMM} shows the SR and FWHM in the K-band obtained with KAPA with infrared tip-tilt (TT) being sensed on IRS29N (which had sufficient SNR for TT correction) at 3 arcsec away from the GC. For all the non-KAPA data in the figure, the infrared TT sensing was done on IRS7 which is 5.5 arcsec away from the GC. So, the performance boost seen for KAPA is a combination of improved LTAO high order performance and reduced anisokinteism owing to IRS29N being closer to the GC (as compared to IRS7).

\subsection{Preliminary sky-coverage estimation}

\begin{figure}[ht]
\centering
\setlength{\fboxsep}{0.15pt}   % Distance between image and border
\setlength{\fboxrule}{0.15pt}  % Border line thickness
\fbox{\includegraphics[width=1.0\textwidth]{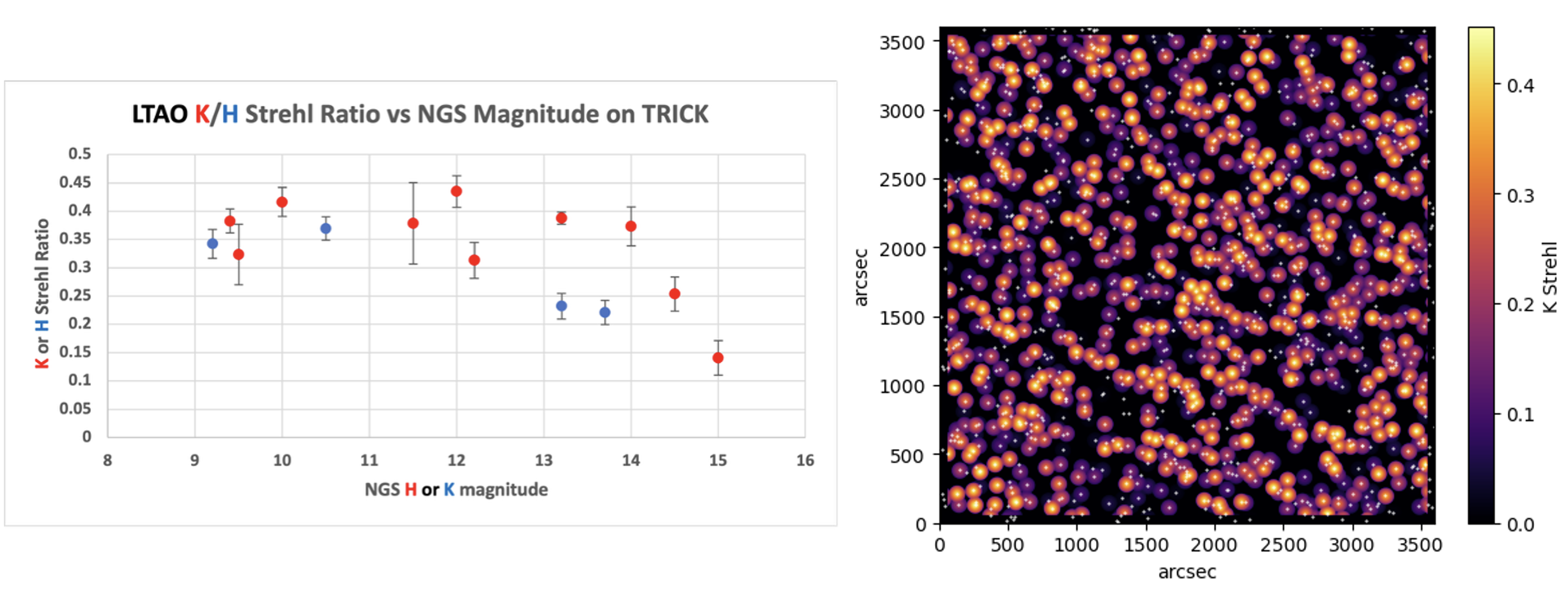}}
\caption{On-axis SR on the OSIRIS Imager in the K and H band when the H and K band (respectively) were being used for TRICK sensing (on the left) and a KAPA Strehl map estimation on a Besancon model generated towards a region located at 60 deg galactic latitude, 90 deg galactic longitude (on the right).}
\label{fig:sky_coverage_inputs}
\end{figure}

\begin{figure}[ht]
\centering
\setlength{\fboxsep}{0.15pt}   % Distance between image and border
\setlength{\fboxrule}{0.15pt}  % Border line thickness
\fbox{\includegraphics[width=1.0\textwidth]{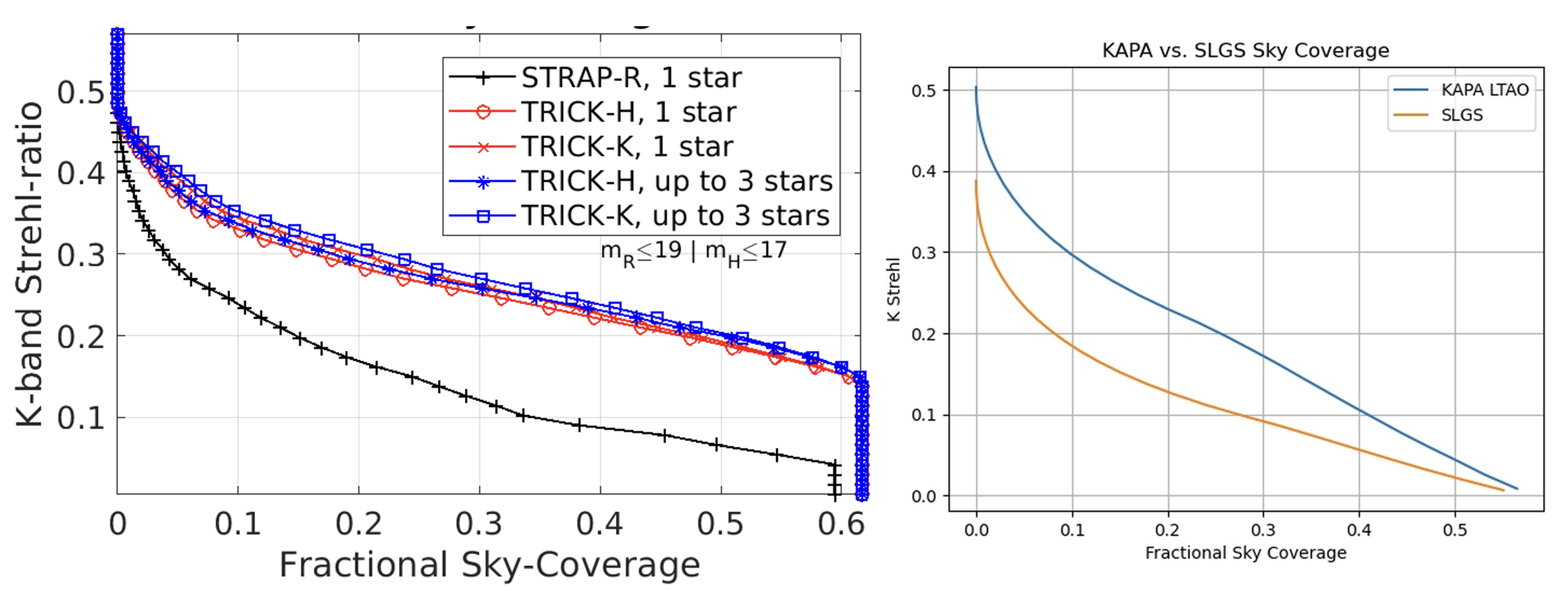}}
\caption{KAPA sky-coverage prediction at the time of design (on the left) and synthesized sky coverage from limited observations (on the right). The data to be compared between the predicted and measured sky coverage are the TRICK-K, 1-star (red, -x-) plot on the left to the KAPA LTAO plot on the right. The SLGS data on the right is shown for showing the sky coverage improvement and does not have a comparable plot on the left.}
\label{fig:sky_coverage_outputs}
\end{figure}

We have performed a very preliminary sky coverage estimation based on the limited data that we have collected towards this measurement. 

The original fractional sky coverage estimation (at the time of setting project requirements) involved the creation of a stellar map (generated by a Besancon model\cite{andersenpredicted}) towards a representative region located at 60 deg galactic latitude, 90 deg galactic longitude with an H magnitude cut-off of 17, followed by the estimation of on-axis SR across the guide stars. In hindsight, two issues with this approach are the lack of consideration of off-axis SR in the stellar gaps and the ambitious correction depth predicted for TRICK (Tilt Removal with IR Compensation, which is the NIR TT sensor for Keck I AO)\cite{rampy2015} sensing upto an H-magnitude of 17. The resulting plot is shown in Fig.~\ref{fig:sky_coverage_outputs} left.

A more realistic estimation is to use the same Besancon model to estimate a representative stellar map towards the same galactic coordinates, but to use the following for sky coverage:
\begin{itemize}
    \item Measure the actual on-axis TRICK sensing depth and SR for sLGS vs LTAO (LTAO shown in Fig.~\ref{fig:sky_coverage_inputs} left and sLGS performance taken as a conservative 90 percent of LTAO over the same magnitudes).
    \item Measure the anisoplanatic angle for both sLGS and LTAO modes.
    \item Combine the above two metrics to model the SR across a Besancon model towards the representative star map at 60 deg galactic latitude, 90 deg galactic longitude (Fig.~\ref{fig:sky_coverage_inputs} right).
    \item Estimate the SR vs sky coverage across the representative field modeled in the above step (Fig.~\ref{fig:sky_coverage_outputs} right).
\end{itemize}
Data used for this approach include:
\begin{itemize}
    \item Images of M79 collected on Jan 31 2026 (UT) point to an effective K-band anisoplanatic angle ($\theta_0$) of 69 $\pm$ 2 arcsec for LTAO as compared to $\sim$ 50 arcsec for sLGS (with the LGS normalized to match the seeing conditions of the LTAO observation).
    \item SR vs TRICK star magnitude (Fig.~\ref{fig:sky_coverage_inputs} left) with different TT stars from M3 collected on May 28 2026 (UT). All data was collected in LTAO mode. 
\end{itemize}

The data to be compared between the predicted and measured sky coverage in Fig.~\ref{fig:sky_coverage_outputs} are the TRICK-K, 1-star (red, -x-) plot on the left to the KAPA LTAO plot on the right. The SLGS data on the right is shown for showing the sky coverage improvement and does not have a comparable plot on the left. It is also to be noted that the predicted performance during design assumes a 4x4 ROI on the TRICK detector while all data was taken with an 8x8 ROI on TRICK. Please note that this estimation is very preliminary (based on one night of TRICK sensing depth and one night of isoplanatic angle estimation) and is subject to change once we collect more data and perform a deeper analysis.

\subsection{Imaging the Arches Cluster}

\begin{figure}[ht]
\centering
\includegraphics[width=1\textwidth]{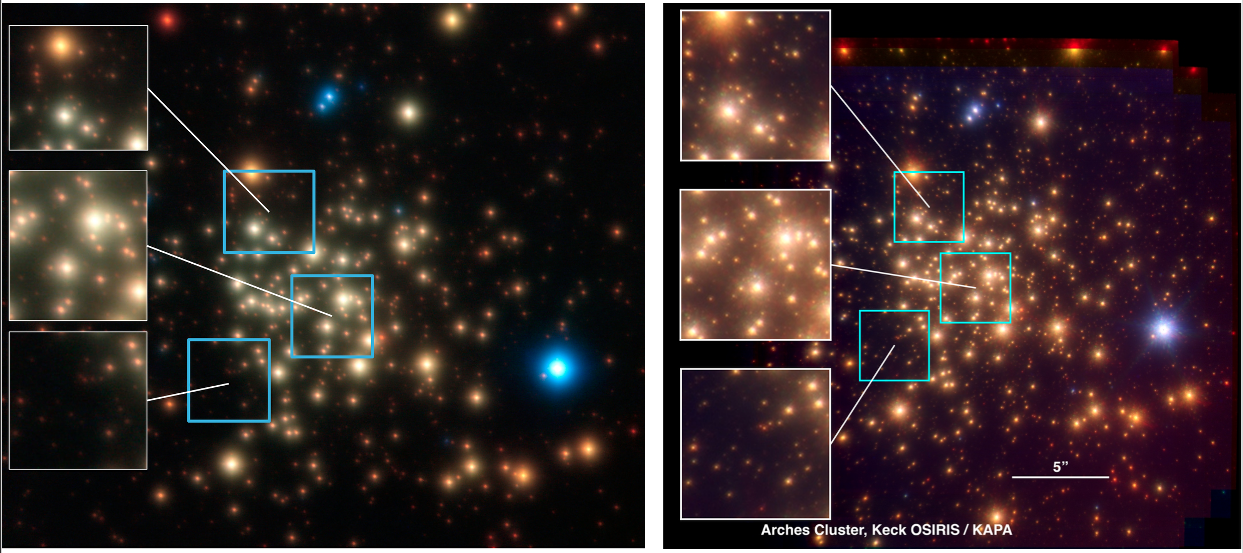}
\caption{A composite J/H/K image of the Arches Cluster from the VLT NACO system (on the left) compared to a similar composite J/H/K image (9 minute integration in each band) from OSIRIS operating in the LTAO mode (on the right).}
\label{fig:Arches}
\end{figure}

The Arches cluster has been used as a near‑infrared laboratory for star formation and massive‑star evolution in an extreme Galactic‑center environment, mainly through high‑resolution imaging, spectroscopy, and astrometry in J/H/K and adjacent narrow bands. Near-infrared science is extensively used to constrain the initial mass function (IMF)\cite{Espinoza2009,Habibi2013}, age, stellar content, winds and metallicity\cite{Clark2018}. The sensitivity of these measurements depend on the depth and resolution that can be obtained on similar crowded fields. Fig.~\ref{fig:Arches} outlines the comparison of a composite J/H/K image of the Arches Cluster from the Very Large Telescope (VLT) Nasmyth AO System (NAOS) Coude Near-Infrared Camera (NACO)\cite{Espinoza2009} with a similar composite J/H/K image (9 minute integration in each band) from OSIRIS operating in the LTAO mode.

For a fair comparison, the comparison of the order of correction between NACO and KAPA are given in Table~\ref{tab:naos_keck_ao}.
\begin{table}[htbp]
\centering
\caption{Comparison of NAOS on the VLT and Keck I AO system}
\label{tab:naos_keck_ao}
\begin{tabular}{lcc}
\hline
Specification & NAOS / VLT & Keck I AO \\
\hline
AO architecture & Single Natural Guide Star AO & Laser tomographic AO (LTAO) \\
Telescope diameter & 8 m & 10 m \\
Wavefront sensor format & 14$\times$14 Shack--Hartmann & 20$\times$20 Shack--Hartmann \\
DM actuators & 185 & 349 \\
Actuator pitch on pupil & 0.57 m & 0.56 m \\
\hline
\end{tabular}
\end{table}

\section{Limitations and operational lessons}

\begin{figure}[ht]
\centering
\includegraphics[width=1\textwidth]{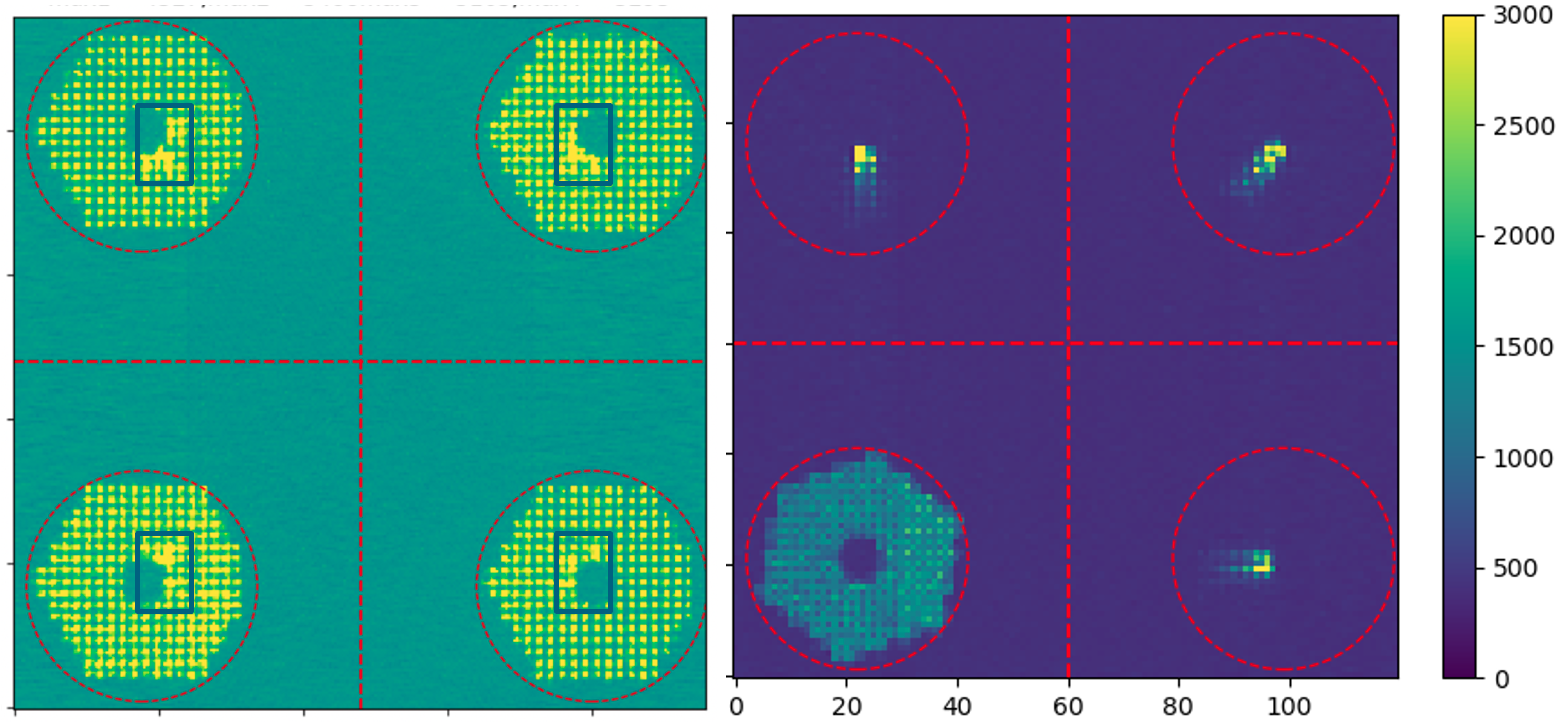}
\caption{Images showing the effect of fratricide on the WFS. The blue boxes in the image on the left shows the fratricide in each pupil from the three remaining laser beams, when all the KAPA loops are closed. Image on the right shows the intensity of the fratricide on the non-illuminated pupils as compared to the laser return on the illuminated pupil.}
\label{fig:fratricide}
\end{figure}

Over the course of commissioning the system, we have come across a few observational limitations:
\begin{itemize}
  \item Thin clouds are known to increase the effect of Rayleigh scattering\cite{vernet1999lgs}. They amplify the effect of laser fratricide (from the neighboring laser) for multi-LGS systems like KAPA. Even in the absence of clouds, the fratricide is substantial (as seen in Fig.~\ref{fig:fratricide}). With the low dynamic range of the OCAM2K sensor, operation of the multi-LGS mode becomes difficult with the fratricide intensity (in the affected subapertures) being almost an order of magnitude higher than the laser return from the sodium layer (in the presence of clouds). At about a cloud extinction of $\sim$1 mag, multi-LGS operation becomes difficult owing to the challenge of setting the camera parameters for preventing overillumination on the fratricide while getting sufficient SNR on the subapertures not affected by fratricide.
  \item The Asterism Generator (inset in Fig.~\ref{fig:kapa-architecture}) rotator is used to derotate the laser asterism on-sky to compensate for the rotation of the asterism by the AO field derotator on the wavefront sensor. It has a range of 280 degrees and does not cover the full rotation range of the AO derotator. So, observations could be interrupted (at most twice in a night) for a rotation procedure to recover range and to re-acquire the target after the procedure (which only takes a minute).
  \item Since we are splitting the 22W laser into four for operating in the multi-LGS mode, we are laser power starved on the wavefront sensor. The result is that we operate at 600 Hz when running in multi-LGS mode as compared to 1.5 kHz for the sLGS mode. There could be few instances of fast seeing where multi-LGS correction may not give a big advantage over sLGS operation (and could be even worse due to the higher bandwidth error associated with fast seeing).
  \item When the WFS optics were upgraded, we inadvertently misaligned the optics and compensated with a misalignment of the focus stage As a result, there is currently ~7 arcsec offset (when pointed at zenith) between the LGS and the NGS conjugate positions of the focus stage. We are planning to realign the WFS beam-train to fix this issue. This is the reason why the peak image quality is off the center of the field in Fig.~\ref{fig:m79_onaxis_srmap} and we lose a bit of performance towards the center of the field. 
\end{itemize}

\section{Path forward}
\label{Sec:path_fw}
The current analysis is incomplete in terms of final validation of performance verification for the KAPA system. The immediate next step is to compare the performance against the level 2 science requirements and for the science verification team to compare performance against level 0 and 1 science requirements. The comparison should include field-dependent performance, stability, throughput, and observing efficiency. We are planning for KAPA to proceed to operational handover in August, marking the transition from engineering validation to routine facility use. Two multi-LGS modes will be offered to the community as part of KAPA:
\begin{itemize}
    \item LTAO-NF (Narrow Field): The tomographic reconstructor is optimized for maximizing the on-axis performance at the center of the asterism.
    \item LTAO-WF (Wide Field): The tomographic reconstructor is optimized for maximizing the on-axis performance across a 3x3 grid with the optimization field across one dimension being [-6.66, 0, 6.66] arcsec for better uniformity of the PSF across the field.
\end{itemize}

KAPA has demonstrated the reduction of focal anisoplanatism and an increase in the effective anisoplanatic angle, and it has been an excellent demonstrator of tomographic AO control on 10 meter class telescopes paving the way for more systems coming online in the near future\cite{heritier2023,Akiyama2020,Noelia2024}. 

The planned addition of the Keck Adaptive Secondary Mirror (KASM)\cite{Phil2024,Proceedings2026Integrated} and the Laser Atmospheric Volumetric Array (LAVA) project will increase the laser power and the actuator count fed to this port and will provide a new wavefront sensor package matched to KASM. KASM and LAVA can upgrade KAPA to provide near--diffraction--limited correction for LGS targets in the Y and J bands for the first time (image quality previously achievable only on the brightest natural guide star targets).

\section{Acknowledgments}
We wish to recognize and acknowledge the very significant cultural role and reverence that the summit of Maunakea has always had within the Native Hawaiian community. We are most fortunate to have the opportunity to conduct observations from this mountain. KAPA was funded by the National Science Foundation (NSF) under the Mid-Scale Innovations Program in Astronomical Sciences (MSIP) Grant Number 1836016. The key science program is funded by the Gordon and Betty Moore Foundation (GMBF).

\appendix
\section{Tomographic reconstructor}
\label{sec:app}
The minimum-variance estimator can be written as the block diagonal of the interaction matrices (\(H_1\) to \(H_4\)):
\begin{equation} \label{eq:Eq1}
G_x = \begin{bmatrix}
    H_1 & 0 & 0 & 0 \\
    0 & H_2 & 0 & 0 \\
    0 & 0 & H_3 & 0 \\
    0 & 0 & 0 & H_4
\end{bmatrix}
\end{equation}
The minimum variance estimate of the wavefront in the direction of the science target \^{x}, is
\begin{equation} \label{eq:Eq2}
\hat{x} = \langle a x^T \rangle \langle x x^T \rangle^{-1} x
\end{equation}
where \textit{x} is the wavefront in the direction of the WFSs and is given by:
\begin{equation} \label{eq:Eq3}
\hat{x} = (\langle x x^T \rangle + G_x^T \langle n n^T \rangle^{-1} G_x)^{-1} G_x^T \langle n n^T \rangle^{-1} s
\end{equation}

where \(\langle x x^T \rangle\) and \(\langle n n^T \rangle\) are the turbulence and noise covariance matrices, respectively. In the KAPA implementation, the noise covariance is approximated by a regularization term (\(\alpha\)) proportional to a diagonal weight matrix \(\alpha^{-1}W\), leading to:
\begin{equation} \label{eq:Eq4}
x_s = (\langle x x^T \rangle + G_x^T \alpha^{-1} W G_x)^{-1} G_x^T \alpha^{-1} W s
\end{equation}
Combining Eq.~\ref{eq:Eq2} and Eq.~\ref{eq:Eq4} gives:
\begin{equation} \label{eq:Eq5}
\hat{a}
=
\alpha^{-1}
\langle a x^{T} \rangle
\langle x x^{T} \rangle^{-1}
\bigl(
  \langle x x^{T} \rangle
  + \alpha^{-1} G_x^{T} W G_x
\bigr)^{-1}
G_x^{T} W s
\end{equation}
which simplifies to:
\begin{equation} \label{eq:Eq6}
\hat{a}
=
\alpha^{-1}
\langle a x^{T} \rangle
\bigl(
  I + \alpha^{-1} G_x^{T} W G_x
\bigr)^{-1}
G_x^{T} W s
\end{equation}

The covariance matrix for the wavefront at two different locations, \(r_1\) and \(r_2\) , and in two arbitrary directions, \(\theta_1\) and \(\theta_2\), is
\begin{equation} \label{eq:Eq7}
C_{\phi}\bigl[(r_1,\theta_1);(r_2,\theta_2)\bigr]
=
c\,(r_0 f_0)^{-5/3}
\sum_{k=1}^{N_l}
\epsilon(k)
\bigl(2\pi f_0 \delta_{\rho}(k)\bigr)^{5/6}
K_{5/6}\!\bigl(2\pi f_0 \delta_{\rho}(k)\bigr)
\end{equation}
where,
\(f_0 = \frac{1}{L_0}\) and \(L_0\) is the outer scale of turbulence and \(K_{5/6}\) is the fractional Bessel function of the second kind of order 5/6. The summation is over \(N_{l}\) turbulence layers, each with fraction \(\epsilon(k)\) of turbulence. 

We define the distance between two wavefronts at altitude layer \(k\), as
\begin{equation}
\delta_{\rho}(k) = \left\lvert \rho_1(k) - \rho_2(k) \right\rvert
\end{equation}
where
\begin{equation}
\rho_i(k) = \left(1 - \frac{h(k)}{z_i}\right) r_i + h(k)\theta_i
\end{equation}

The altitude of the guide star is \(z_i\) and the altitude of the turbulence layer is \(h(k)\) in the above equation. 

\(c\) is a constant given by:

\begin{equation}
c =
\left(\frac{24}{5}\Gamma\!\left(\frac{6}{5}\right)\right)^{5/6}
\frac{\Gamma\!\left(\frac{11}{6}\right)}{2^{5/6}\pi^{8/3}}
\end{equation}

To compute the matrices, we adopt a representative turbulence profile for Maunakea (Table~\ref{tab:covariance_parameters}). The reconstructor is not very sensitive to the modeled turbulence profile but has a small chance of impacting performance if the actual atmospheric profile is very different from this model.

\begin{table}[htbp]
\centering
\begin{tabular}{|l|l|l|}
\hline
Quantity & Value & Units \\
\hline
$L_0$ & 30 & m \\
$r_0$ & 0.15 & m \\
Layer altitudes & $[0, 500, 1000, 2000, 4000, 8000, 16000]$ & m \\
Turbulence fraction & $[0.4557, 0.1295, 0.0442, 0.0506, 0.1167, 0.0926, 0.1107]$ & - \\
Guide star locations & $[[5.5, 5.5], [5.5, -5.5], [-5.5, -5.5], [-5.5, 5.5]]$ & arcsec \\
\hline
\end{tabular}
\caption{Parameters used in the covariance matrix calculations.}
\label{tab:covariance_parameters}
\end{table}

To calculate the covariance matrix $\langle a x^{T} \rangle$ between the wavefront at the WFSs
and the wavefront at the science target, we assume that we are trying to optimize for a science target
at the optical axis. This choice can change based on whether we are using LTAO-WF or LTAO-NF as described in Section~\ref{Sec:path_fw}.

\section{Pseudo open loop control (POLC)}
\label{Sec:appendix_polc}
The standard controller in sLGS AO is a leaky integrator operating in closed-loop:
\begin{equation}
y[n]
=
-b_1 \, y[n-1]
+ a_0 \, u[n]
\end{equation}
where $y[n]$ is the mirror command at time n, $u[n]$ is the residual at time $n$, $b_1$ is the leak factor and is typically -0.99 and
$a_0$ is the loop gain. The residual reconstruction is given by:
\begin{equation}
u = R s
\end{equation}
where $R$ is the sLGS reconstructor and s is the slope vector. In POLC, we reconstruct the reconstructed open loop centroids (from the DM shape $y[n-1]$ and the interaction matrix $H$) as follows:
\begin{equation}
u[n]
=
R \bigl( s[n] + H y[n-1] \bigr)
- y[n-1]
\end{equation}
This can be rearranged into:
\begin{equation}
u[n]
=
R s[n]
+ (R H - I) y[n-1]
\end{equation}
and hence
\begin{equation}
u[n]
=
R s[n]
+ D y[n-1]
\end{equation}
where $D=RH-I$. The sign convention at Keck saves the reconstructor with a negative sign as compared to the sign of the interaction matrix. Hence the 'POLC matrix' $D$ is given by:
\begin{equation}
D = -\bigl( I + R H \bigr)
\end{equation}
 
\bibliographystyle{spiebib}
\bibliography{report}

\end{document}